\documentclass{optica-article}
\journal{opticajournal} 

\articletype{Research Article}
\usepackage{bm}

\begin{document}

\title{High-background X-ray single particle imaging enabled by holographic enhancement with 2D crystals}

\author{Abhishek Mall,\authormark{1,2} Zhou Shen,\authormark{1,2} and Kartik Ayyer\authormark{1,2,3,*}}

\address{\authormark{1}Max Planck Institute for the Structure and Dynamics of Matter, 22761 Hamburg, Germany\\
\authormark{2}Center for Free-Electron Laser Science, 22761 Hamburg, Germany\\
\authormark{3}The Hamburg Center for Ultrafast Imaging, Universität Hamburg, 22761 Hamburg, Germany\\
}

\email{\authormark{*}kartik.ayyer@mpsd.mpg.de} 

\begin{abstract*} 
X-ray single particle imaging (SPI) has offered the potential to visualize structures of biomolecules at near-atomic resolution. However, state-of-the-art structures at X-ray free electron lasers (XFELs) are limited to moderate resolution, primarily due to background scattering. We computationally explore a modified SPI technique based on holographic enhancement from a strongly scattering 2D crystal lattice placed near the object. The Bragg peaks from the crystal enable structure retrieval even for background levels up to 10$^{5}$ times higher than the object signal. This method could enable SPI at more widely accessible synchrotron sources, where even detection of objects before radiation damage is nearly impossible currently, supports practical fixed-target sample delivery, and enables high-resolution imaging under near-native conditions. Numerical simulations with a custom reconstruction algorithm to recover the latent parameters show the potential to improve the achievable resolution while also expanding the accessibility to the technique.
\end{abstract*}

\section{Introduction}

Biomolecules, such as viruses and large protein complexes, play an essential role in understanding fundamental life processes, including their functionality and survival mechanisms~\cite{nasir2015phylogenomic,srihari2015methods}. The structures of these biological complexes largely determine their functional properties, making the imaging of their architecture and  dynamics an essential aspect in structural biology~\cite{dean2024illuminating}. 

X-ray single particle imaging (SPI) is a technique for visualizing the structure and ultrafast dynamics of nanoscale entities, such as viruses and inorganic nanoparticles, in near-native environments~\cite{ekeberg2015three,mall2024observation,ayyer20203d,shen2025direct}. Traditionally, X-ray crystallography has been used to determine atomic resolution biomolecular structures, but this relies on being able to crystallize the molecules, which leads to inherent limitations~\cite{seibert2011single}. Cryo-electron microscopy has rapidly emerged as a viable alternative, where the structure is reconstructed from a large number of single particle in-line holograms with particles in random orientations. Due to the ability to measure at room-temperature and the high time-resolution possible, X-ray SPI remains of interest as a complementary method.

SPI has shown its greatest successes at X-ray free-electron laser sources (XFELs), which provide ultrabright, ultrafast pulses ideally suited for imaging weakly scattering biological particles~\cite{spence2017xfels,giewekemeyer2019experimental,sobolev2020megahertz} and achieving 3D structural reconstruction with nanometer-scale resolution~\cite{ayyer20203d,PhysRevE.80.026705,ayyer2016dragonfly}. It has also been applied to disentangle conformational heterogeneity and track ultrafast dynamics~\cite{mall2024observation,shen2024resolving,zhuang2022unsupervised,dold2025melting,shen2025direct}.
However, XFEL access is limited, and optimizing sample delivery to minimize experimental and instrumental background remains a major challenge~\cite{bielecki2019electrospray,zhao2019guide,yenupuri2024helium,you2024impact}. Synchrotrons--being more widely available and increasingly brilliant~\cite{helliwell1998synchrotron,neutze2012time,khubbutdinov2019coherence}--offer an attractive alternative for routine SPI. 

While aerosol-based sample delivery minimizes background at XFELs~\cite{bielecki2019electrospray}, synchrotrons have too low a peak intensity to allow diffraction measurements from fast moving particles. Fixed-target approaches, though more practical at synchrotrons, introduce additional substrate scattering that degrades data quality~\cite{nam2016fixed,hunter2014fixed,seuring2018femtosecond}. Moreover, current X-ray imaging at synchrotrons is largely limited to larger particles that scatter strongly enough to produce detectable patterns~\cite{chushkin2025prospects,sun2018current}. These challenges, compounded by radiation damage, mean that SPI at synchrotrons has been considered essentially impossible.

\begin{figure}
\centering
\includegraphics[width=0.95\textwidth]{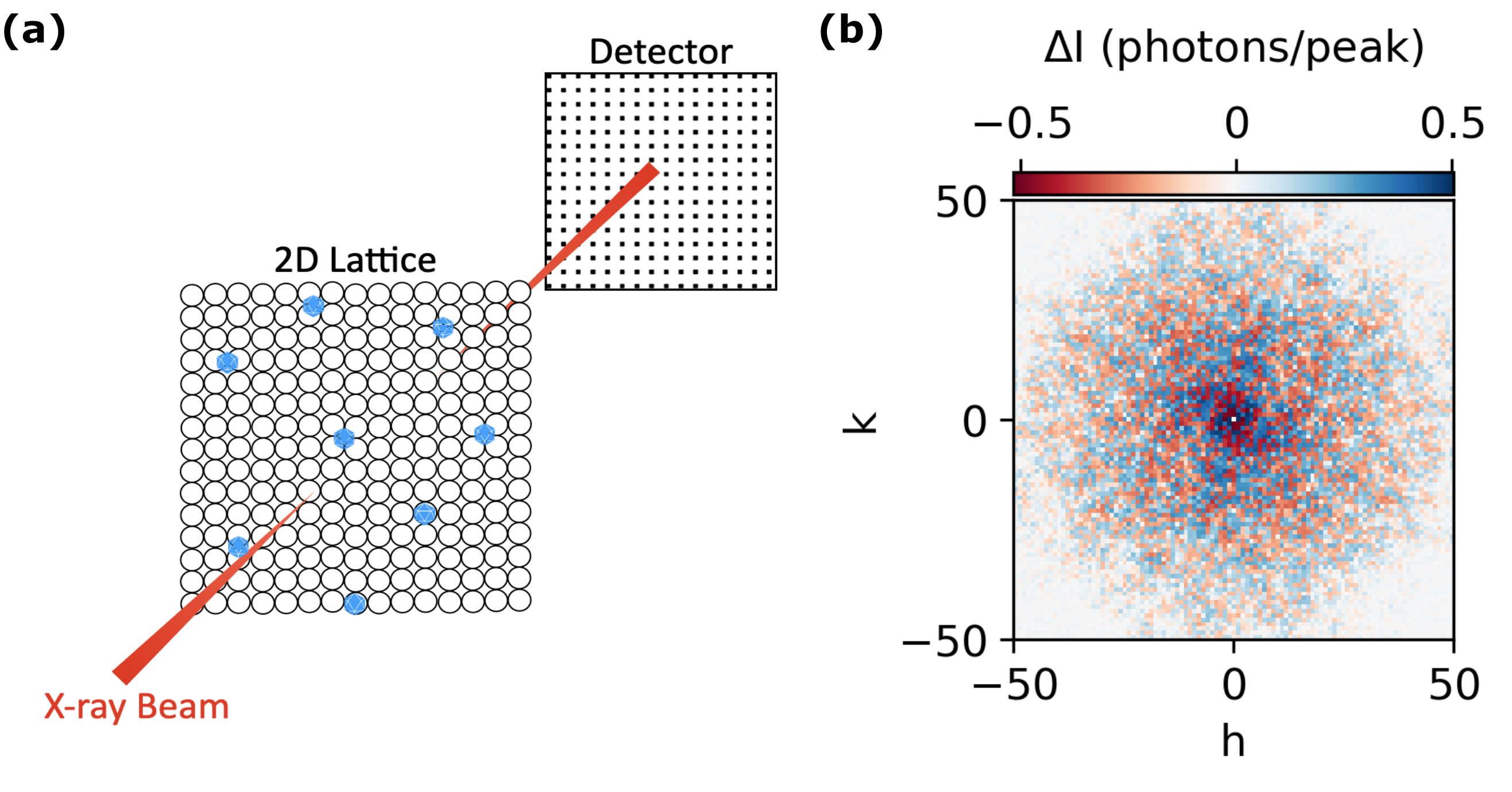}
\caption{\textbf{Holographic SPI. (a)} Schematic of the holographic SPI setup. A 2D crystal lattice serves as the \textit{reference}, while ribosome particles (blue) are the target objects. The lattice is positioned on one side of the substrate, with target particles randomly distributed on the opposite side. The measured far-field diffraction pattern on the detector results from the coherent sum of Bragg peaks from several illuminated unit cells of the lattice and the scattering signal from a single target particle. 
\textbf{(b)} Far-field intensity modulation shown on a logarithmic scale. These modulations reflect Bragg peaks perturbed by the target object signal within the illumination area. The modulation, $\Delta I$, is defined as the difference between the intensity from the lattice alone, $I = |F_{\text{L}}(\textbf{q})|^2$  and the intensity with both the lattice and the target object $I = |F_{\text{L}}(\textbf{q}) + \varphi \cdot F_o(R^\top\textbf{q}) e^{i 2\pi \textbf{q} \cdot \textbf{s}}|^2$, for a given translation shift within unit cell, in-plane orientation, and incident fluence.}
  \label{fig:lattice_spi}
\end{figure}

Here, we propose an experimental strategy to enable high-resolution SPI at synchrotrons by employing a strongly scattering 2D crystal lattice as a holographic reference in a fixed-target setup. This approach adapts the ``reference-enhanced'' holographic SPI approach, initially proposed for XFELs, in which a reference scatterer like a gold nanoparticle was used, and the holographic cross-term led to improved performance~\cite{ayyer2020reference,mall2023holographic}. In our adaptation, a 2D crystal lattice is fabricated on one side of a thin substrate, producing sharp Bragg peaks upon X-ray illumination, while target biomolecules are randomly dispersed on the opposite side, as illustrated in Fig.~\ref{fig:lattice_spi}a~\cite{vogel2015advances,ye2017monolayer}. Interference between the lattice diffraction and the scattering from the target object leads to modulations in the peak intensities (Fig.~\ref{fig:lattice_spi}b). In addition, we will generate continuous diffraction from the single target object, but that is assumed to be drowned in the background. This holographic interference significantly enhance the signal-to-background ratio (SBR) to such an extent that imaging is possible even when the background noise exceeds the object signal by up to five orders of magnitude. A tailored reconstruction algorithm iteratively determines the structure of object, orientation, translation shift within the unit cell, and incident beam fluence, eliminating the need for additional support constraints and minimizing modeling assumptions~\cite{mall2023holographic}.

Simulations using a 70S ribosome model biomolecular complex (PDB:7NHM)~\cite{crowe2021structural} demonstrate that this lattice-based holographic approach can resolve target structures under conditions of high background scattering and variable beam exposure. This framework broadens the scope of SPI, enabling routine, high-resolution imaging of diverse biomolecular assemblies at synchrotron facilities and establishing a path toward efficient, standardized structural studies.

\section{Problem Formulation}
This section describes the framework used to simulate holographic single particle imaging a 2D crystal lattice reference scatterer. In this fixed-target configuration, the lattice (either patterned onto a chip or self-assembled as a colloidal crystal~\cite{ye2017monolayer}) is placed on one side of a thin substrate while target object particles are randomly dispersed on the opposite side, as shown in the schematic in Fig.\ref{fig:lattice_spi}a. This configuration generates structured interference patterns with Bragg peaks on the detector that encode information from both the lattice and the target object.

To model this setup through 2D numerical simulations considering experimental conditions. The unit cell electron density of a square lattice, denoted as $\rho_L(\mathbf{r})$, generates sharp Bragg peaks and $\rho_o(\mathbf{r})$, represents the projected electron density of the \textit{ribosome} target object. The total scattering contrast, incorporating a rotated and translated target object within a unit cell of the lattice is then expressed as 
\begin{equation}
\rho(\mathbf{r}) = \rho_L(\mathbf{r}) + \rho_o(\bm{R}\cdot\mathbf{r} - \mathbf{s}),
\label{eq:edens}
\end{equation}
where $\bm{R}$ corresponds to rotation operator for the in-plane orientation ($\theta$) uniformly sampled from $[0, 2\pi)$. The vector $\mathbf{s}$ denotes the translational shift of the target object within the unit cell along $x$- and $y-$ direction, uniformly sampled from $[0, 1)$ expressed in fractional coordinates. Due to the periodicity of the lattice, shifts can be wrapped to the $[0,1)$ domain. And if the unit cell is not a square, we will operate in the space of the basis vectors and convert to true $x-y$ coordinates with an affine transformation in the end. For this study, we assume the unit cell is a square. 

The corresponding intensity measured on the detector incorporating a slowly varying background $B^2(\mathbf{q})$, is given by 
\begin{equation}
I(\mathbf{q}) = \left| F_L(\mathbf{q}) + \varphi  F_o(\bm{R}\cdot\mathbf{q}) e^{i 2\pi \mathbf{q} \cdot \mathbf{s}} \right|^2 + B^2(\mathbf{q}),
\label{eq:intens}
\end{equation}
where $F_L = \zeta \cdot \mathcal{F}[\rho_L]$ is the scaled Fourier transform of the lattice, $F_o = \mathcal{F}[\rho_o]$ is the Fourier transform of the target object, and $\varphi$ is the incident fluence on the target object. The strong scattering from the lattice serves as a holographic reference, facilitating signal extraction of target object and translation shifts estimation. It's relative scattering strength compared to the target is controlled by $\zeta$ factor (discussed in Section~\ref{subsec:lattice_strength}) and plays a critical role in signal retrieval, noise tolerance, and the optimization of reference design.

The phase term, $e^{i 2\pi \mathbf{q} \cdot \mathbf{s}}$, encodes the translational shift of the target object within the unit cell, while the background term, $B(\mathbf{q})$, accounts for contributions from the substrate, noise, and beamline scattering. Consequently, the observed intensity, incorporating photon counting noise is: 
\begin{equation}
I_{\text{obs}}(\mathbf{q}) = \frac{\text{Poisson}(I(\mathbf{q}) \cdot t)}{t} - B^2(\mathbf{q}),
\label{eq:Iobs}
\end{equation}
where $\text{Poisson}(\lambda)$ denotes the generation of a Poisson distributed random variable with mean $\lambda$ and $t$ is the exposure time in seconds. In practical experiments, the background $B^2(\mathbf{q})$ can be estimated from dark measurements -- taken without the reference or target particles -- or by fitting smooth functions to detector regions away from the Bragg peaks. The 2D lattice reference can be experimentally characterized using approaches similar to probe characterization methods commonly employed in ptychography~\cite{thibault2009probe}.

Eq.~\ref{eq:intens} contains a cross term that plays a key role in determining the SBR. As shown in~\cite{ayyer2020reference}, this term includes a cosine component whose amplitude carries important information about translational shifts. A strong reference improves the robustness of the signal: although it increases the overall noise level, it also makes the signal less sensitive to background. The amplitudes of the cross term's fluctuations are crucial, since if they are large enough, the relative positions of the target can be extracted from the diffraction patterns. These amplitudes also affect the feasibility of reconstructing the complex function $F_o(\mathbf{q})$.

\subsection{Effect of Scattering Strength of Lattice}
\label{subsec:lattice_strength}

\begin{figure}
\centering
\includegraphics[width=0.55\textwidth]{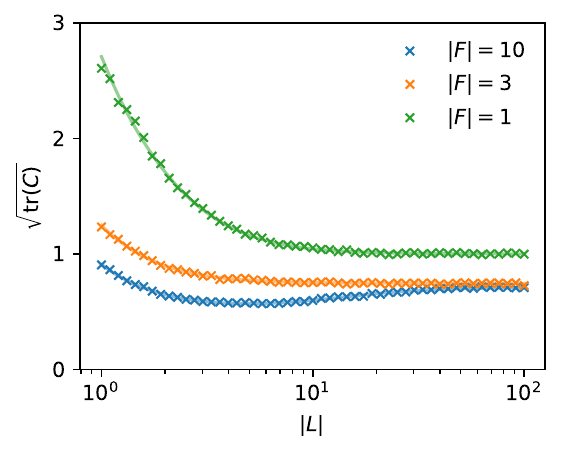}
\caption{\textbf{Scattering Strength of Lattice.} The plot illustrates the uncertainty in estimating the scattering amplitude of $F_o(\mathbf{q})$, quantified by $\sqrt{\mathrm{tr}(C)}$ (the square root of the combined amplitude and phase variances of $F_o(\mathbf{q})$) as a function of the scattering strength of the 2D crystal lattice, with a fixed background $B^2(\mathbf{q})$. Results are shown for three different scattering amplitudes of $F_o(\mathbf{q})$. Solid lines depict analytical estimates based on the inverse curvature of the objective function, while dots represent uncertainties obtained from simulated data through curve fitting.}
\label{fig:lattice_strength}
\end{figure}

One of the main experimental design questions regards the nature of the 2D crystal lattice. In order to achieve this, we study the influence of lattice strength on the accuracy of retrieving $F_o$ at a particular $\mathbf{q}$ when the ground truth of the latent variables is assumed to be known. Specifically, in Eq.~\ref{eq:intens}, all orientations are assumed to be identical, $\varphi = 1$, and the phases $\phi = 2 \pi \mathbf{q} \cdot \mathbf{s}$ are known and uniformly distributed over $\mathcal{R}$. For a fixed $\mathbf{q}$, we examine the accuracy of fitting at $F=F_o(\mathbf{q})$ from a joint distribution $(\phi, P)$ where the photon count $P$ follows a Poisson distribution, $\text{Poisson}[I(F; L, B, \phi)]$, with the simplified intensity model defined as: 
\begin{equation}
I(F'; L, B, \phi) = \left| L + F' e^{i\phi} \right|^2 + B^2\text{,}
\label{eq:Isimple}
\end{equation}
and  $L = F_L(\mathbf{q})$.
We obtain $F$ by performing a least-squares fit by minimizing the following $L_2$ loss function:
\begin{equation}
T(F'=a e^{i r})=\mathbb{E}_{\phi} \mathbb{E}_{P\sim \text{Poisson}[I(F)]} \frac{1}{2}\bigl[I(F'; L, B, \phi) - P\bigr]^2
\label{eq:target}
\end{equation}
The Hessian matrix $H$ of $T$ evaluated around the ground truth $F'=F$ is
\begin{equation}
H 
= \begin{pmatrix}
   \partial_a  \partial_a&  \partial_a  \partial_r\\
   \partial_a  \partial_r&  \partial_r  \partial_r
\end{pmatrix}
T(F'=F)
= 2 \begin{pmatrix}
    |L|^2 + 2 |F|^2 & 0 \\
    0 & |L|^2 |F|^2 \\
\end{pmatrix}
\text{, } 
\end{equation}
with the corresponding variance $\sigma^2 = 2 T(F'=F) = |L|^2 +|F|^2+B^2$.
The total uncertainty in the fitted result is then given by:
\begin{equation}
    \sqrt{\mathrm{tr}(C)} = \Biggl[\frac{|L|^2+|F|^2+ B^2}{2} \biggl( \frac{1}{|L|^2 + 2 |F|^2} + \frac{1}{|L|^2 |F|^2}\biggr) \Biggr]^{\frac12}\text{,}
\end{equation}
where $C = \sigma^2 H^{-1}$ is the covariance matrix.
Figure~\ref{fig:lattice_strength} illustrates distinct regimes depending on the relative magnitudes of 
$B$ and $|F|$. Each point in the plot represents a numerically estimated uncertainty based on 10,000 samples of $(\phi, P)$.




For low scattering amplitudes $|F|^2 \lesssim B^2$, the uncertainty decreases monotonically with increasing lattice strength. Thus, stronger scattering lattices improve the accuracy of recovering $F$. For higher scattering amplitudes $|F|^2 \gg B^2$, an optimal lattice strength emerges around $|L| \approx \sqrt{F}$, beyond which further increases in $|L|$ reduce accuracy. This suggests that, for large scattering amplitudes of $F$, matching the lattice strength to the background intensity is advantageous.

Practically, $\zeta$ should be experimentally optimized to ensure lattice scattering strength matches or exceeds background noise intensity, maximizing reconstruction fidelity. For our simulation we keep the scattering strength of lattice as large as  possible by allocating $\zeta$ to be high. This is also suggested in \cite{marchesini2008massively}, where uniformly redundant arrays (URAs) serve as structured references with high and spatially uniform scattering strength. These arrays provide a near-constant scattering strength up to high  frequencies, significantly enhancing retrieval.

\subsection{Reconstruction Algorithm}
We begin by defining the model intensity, which the reconstruction algorithm fits to the measured diffraction data
\begin{equation}
I_{\mathrm{calc}}(\mathbf{q})
=  \bigl|\,F_L(\mathbf{q})
+ \varphi \,F_o(\bm{R}\cdot\mathbf{q})\,e^{i2\pi\,\mathbf{q}\cdot\mathbf{s}}\bigr|^2
\label{eq:I_model}
\end{equation}
The goal of our reconstruction is to estimate four unknown parameters from each diffraction pattern: (a) the translational shift of the object, denoted by the 2D vector $\mathbf{s}$, representing the position of the target object within the unit cell along the $x$- and $y$-directions, (b) the incident fluence, $\varphi$, which quantifies the total number of photons illuminating the particle and effectively captures the degree of overlap between the particle and the beam, (c) the in-plane orientation of the object, $\theta$, and (d) the complex-valued Fourier transform of the target object, $F_o(\mathbf{q})$, which upon inverse Fourier transform, yields the projected real-space electron density. Each diffraction frame is modeled as a noisy intensity measurement, as described in Eq.~\ref{eq:Iobs}, and is associated with a randomly sampled orientation $\theta$, translation $\mathbf{s}$, and fluence $\varphi$. The exposure time, denoted $t$ (in seconds), determines the total photon flux and thereby influences the signal-to-background ratio. Diffraction intensities are recorded on an $N\times N$ detector grid in reciprocal space. The corresponding real-space grid has dimensions 101$\times$101 pixels, with each pixel representing 10\text{\AA} with a total unit cell size of 100 nm.

Given a flux of $10^{11}$ photons/$\mathrm{\mu}$m$^2$/s, simulations with the\textit{Dragonfly} software~\cite{ayyer2016dragonfly} report $3.56 \times 10^5$~photons/sr at a scattering angle corresponding to a resolution of 100~nm after 1 second of exposure. A 1~$\mathrm{\mu}$m beam at 8~keV photon energy subtends a solid angle of $2.4\times10^{-8}$~sr at the position of a Bragg peak. Accordingly, the expected number of photons within such a Bragg peak is $8.54 \times 10^{-3}$.

To reconstruct \textbf{s}, $\varphi$, and $F_o(\mathbf{q})$, the algorithm minimizes the discrepancy between predicted and measured diffraction intensities. Specifically, at each iteration, we minimize 
\begin{equation}
\mathcal{E} = \sum_{\mathit{j}} \sum_{\mathbf{q}} \left[ I_{\text{obs}, j}(\mathbf{q}) - I_{\text{calc}}(\mathbf{q}|\textbf{s}_j , \varphi_j, \theta_j, F_o) \right]^2
\label{eq:err}
\end{equation}
where $I_{\text{obs}, j}(\mathbf{q})$ represents the observed intensity of frame, $j$, and $I_{\text{calc}}(\mathbf{q})$ is the predicted intensity of $j$. This is carried out in an iterative cycle of three key updates through a systematic grid search over the unknown parameters, inspired by the pattern search method in ~\cite{mall2023holographic}. The search proceeds in two stages: an initial coarse exploration over a broad parameter space, followed by a fine search around the best coarse estimate to refine the solution. While one can perform this 2D optimization in the complex plane with many methods, we found that with a grid search approach, it could be efficiently parallelized on the graphical processing units (GPUs) and secondly it was quite robust in high background limit.
 
Firstly, the algorithm begins by initializing $F_o(\mathbf{q})$, either with random values or using a circular object estimate sized similarly to the target object, where size estimates may be derived from SAXS studies~\cite{li2016small}. For the simulations presented here, $F_o(\mathbf{q})$ is initialized randomly and in-plane orientation $\theta$. To estimate the in-plane translation shifts \textbf{s} and incident fluence $\varphi$, a coarse grid search samples possible values across a broad range and identifies the best candidate by comparing the measured and predicted intensities. A finer “zoomed-in” search then refines this estimate. In practice, each frame has its own shift and fluence, so these computations are repeated frame by frame.

Secondly, given the updated \textbf{s} and $\varphi$, we next solve for the in-plane rotation angle $\theta$. A coarse angular sweep from 0 to 2$\pi$ locates a region of minimal error, followed by progressively narrower bracket searches around this best angle. We gradually reduce the step size until no significant further improvement is found.

Lastly, with \textbf{s}, $\varphi$ and $\theta$ fixed, we update $F_o(\mathbf{q})$ on each pixel $\mathbf{q}$. The unknown amplitude is treated as a complex number, and we perform a coarse-fine grid search on the real and imaginary part of F separately to minimize the squared error across all frames. This entire three-step process constitutes one iteration. The algorithm then repeats these parameter updates for multiple iterations (e.g., 100) until convergence or a preset limit is reached.
 
This coarse-to-fine search approach, although computationally intensive for large experimental datasets, is made feasible by leveraging parallelized computation on graphical processing units (GPUs). Future experimental implementations would benefit significantly from adaptive optimization algorithms to ensure efficient processing of large datasets.

\section{Results}

\begin{figure}
\centering
\includegraphics[width=0.85\textwidth]{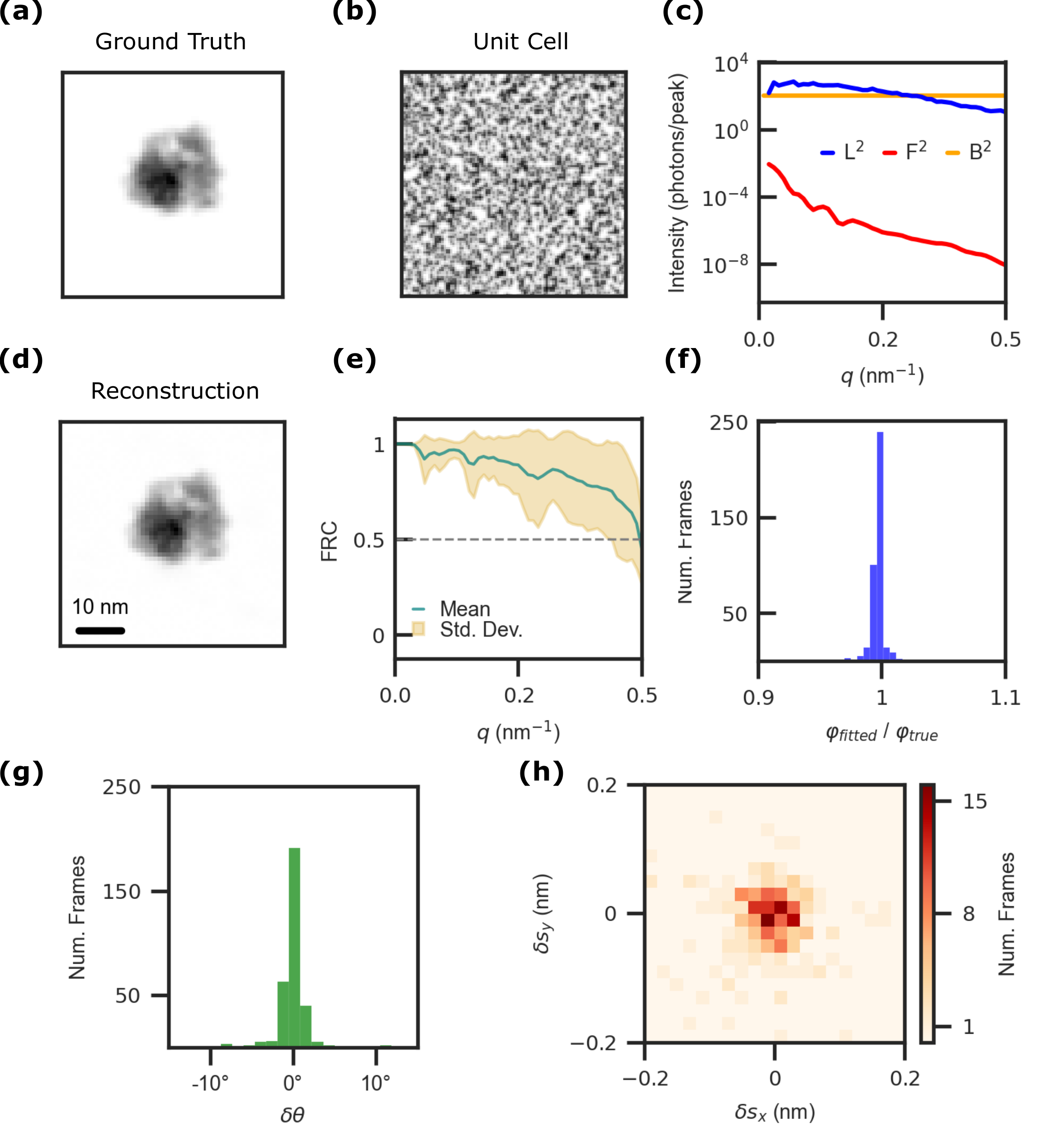}\\
\caption{\textbf{Performance}. 
\textbf{(a)} Ground truth 2D projection of the ribosome electron density. 
\textbf{(b)} Unit cell of the 2D crystal lattice, modeled as random noise. \textbf{(c)} Scattering intensity as a function of $q$ for the 2D lattice ($blue$), ribosome target object ($red$), and background ($yellow$). The background exceeds the signal of ribosome by more than $10^5$, while the lattice intensity is comparable to or greater than the background. \textbf{(d)} Reconstructed electron density of the ribosome (scale bar: 10 nm). \textbf{(e)} Fourier ring correlation (FRC) between the reconstruction and the ground truth $F_o(q)$. Green lines indicate FRC values averaged over 8 random-start reconstructions; the shaded region represents the standard deviation across runs.  
\textbf{(f)} Ratio of estimated to true incident fluence, $\varphi$. 
\textbf{(g)} Distribution of estimation errors for in-plane orientation angle $\theta$ across random runs. 
\textbf{(h)} Distribution of translational shift errors in the x and y directions for the target object within the unit cell (unit cell size: 100 nm). The performance is evaluated for exposure time \textit{t}=1.}
\label{fig:performance}
\end{figure}

\begin{figure}
\centering
\includegraphics[width=0.9\textwidth]{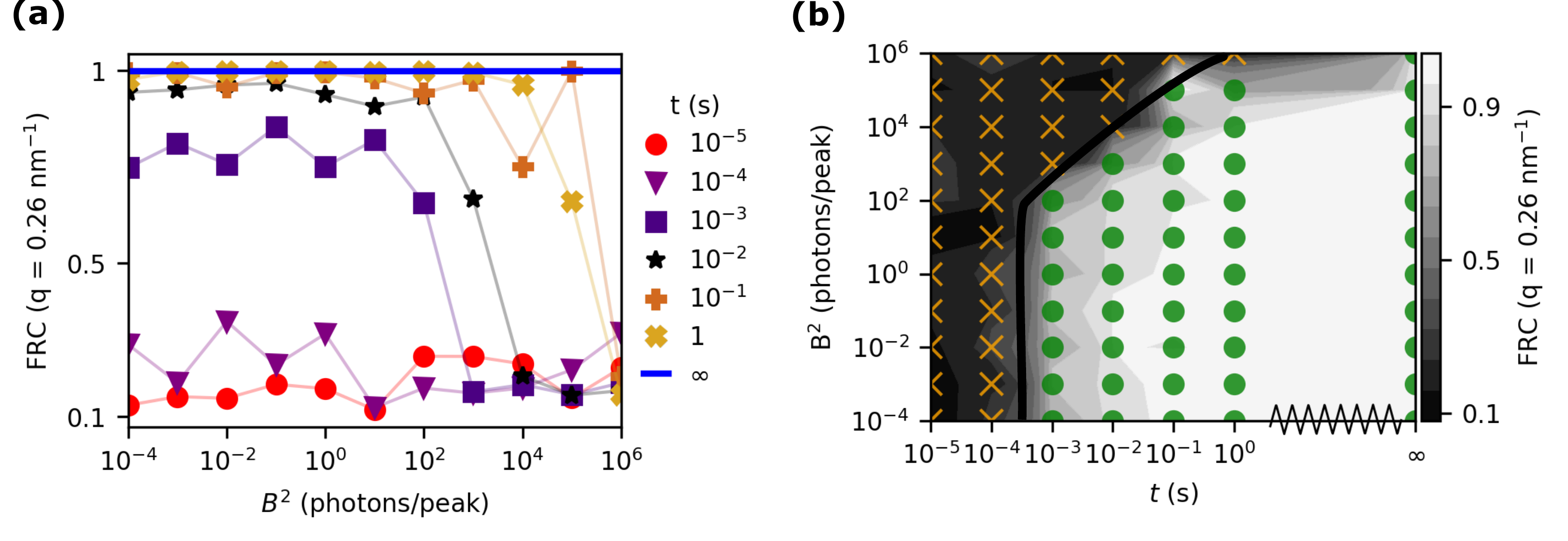} \\
\caption{\textbf{Evaluation of Reconstructions.} 
\textbf{(a)} Fourier ring correlation (FRC) values at $q = 0.26\,\text{nm}^{-1}$ \textit{vs} background $B^2$, for various exposure times ($t$). Each data point represents the maximum FRC value from eight random-start reconstruction runs. The horizontal solid line at FRC = 1 denotes the theoretical limit corresponding to infinite exposure time, $t \to \infty$. 
\textbf{(b)} Contour plot of FRC values (color scale) as a function of $B^2$ and $t$, both on logarithmic scales. Green circles indicate regions where FRC $\geq$ 0.5, orange crosses mark regions with 0 < \text{FRC} < 0.5, and the black line traces the threshold boundary at FRC = 0.5. The sawtooth pattern along the bottom edge represents the approach to infinite exposure time $t \to \infty$ where FRC reaches 1.}
\label{fig:bg}
\end{figure}

We assess the performance of the reconstruction algorithm within the holographic SPI framework using a 2D crystal lattice. This evaluation focuses on both the quality of the reconstructed image and the error distribution in the estimated latent parameters. Figure~\ref{fig:performance} presents a comprehensive analysis, including signal strength comparisons, reconstruction accuracy, and the precision of parameter estimation for a single experimental condition.

Figures~\ref{fig:performance}a–b display the ground truth 2D projection of the ribosome electron density and the unit cell of the 2D crystal lattice, modeled as random noise. Figure~\ref{fig:performance}c shows the scattering strength as a function of $\mathbf{q}$ for the 2D lattice ($L = F_L(\mathbf{q})$), the target object ($F = F_o(\mathbf{q})$), and the background ($B(\mathbf{q})$). Notably, the background intensity exceeds that of the target object by many orders of magnitude up to $10^5$, posing a major challenge for conventional SPI techniques, where weak biological signals are often obscured by overwhelming background noise, making structural reconstruction nearly impossible. In contrast, within the conjugated system used here, the lattice signal remains comparable to or even stronger than the background at low $\mathbf{q}$ values, ensuring the visibility of Bragg peaks. These peaks are modulated by the presence of the target object, allowing detection even in high-noise conditions. This approach relies on the assumption that both the lattice signal and background can be accurately characterized.

Figure~\ref{fig:performance}d presents the reconstructed electron density map of the ribosome, demonstrating that the algorithm can successfully recover fine structural details despite the weak and noisy diffraction signal shown in Fig.~\ref{fig:performance}c, even in the absence of support constraints or any prior information about the target object. The reconstruction accurately captures the major features of the target object, as shown in Fig.~\ref{fig:performance}e using the Fourier ring correlation (FRC) metric. The solid curve shows the mean FRC across eight independent runs with randomized initializations and diffraction datasets, while the error bars represent the standard deviation at each \textit{q}. The FRC consistently exceeds the 0.5 threshold (the half-resolution criterion) across the entire \textit{q} range, indicating reliable and reproducible reconstructions. The narrow spread of the error bars further highlights the algorithm's robustness to noise and initialization variability. These results correspond to a diffraction dataset with an exposure time of $t = 1\,\mathrm{s}$.

Panels f–h of Figure~\ref{fig:performance} assess the accuracy and precision of the reconstructed latent parameters: incident fluence ($\varphi$), in-plane orientation ($\theta$), and translational shifts (\textbf{s}). Results are aggregated across all diffraction frames from the eight independent reconstruction runs. In particular, Fig.~\ref{fig:performance}h presents a 2D histogram of translational shift errors within the unit cell. The majority of estimates fall within $\pm0.2$ nm of the true position in both $x$ and $y$ directions, which is significantly better than the highest resolution in the dataset of 2~nm. Together, these results highlight the robustness and effectiveness of the method under realistic noise conditions, showing strong resilience to background interference, accurate estimation of experimental parameters, and consistent structural recovery across multiple runs.

To assess the conditions under which our reconstruction algorithm remains effective, we systematically evaluated its performance across a range of background levels, $B^2$ and exposure time, $t$. Figure~\ref{fig:bg}a shows the FRC value at \textit{q} = 0.26 nm$^{-1}$ as a function of these parameters. At low exposure times, the weak scattering from both the lattice and the target object fails to generate reliable interference patterns, resulting in reconstruction failure. Figure~\ref{fig:bg}b presents a contour plot of the FRC values at \textit{q} = 0.26 nm$^{-1}$, mapped over logarithmic axes of 
$B^2$ and $t$. The plot reveals a distinct transition boundary (black line) where the FRC drops below 0.5, marking the threshold for reliable reconstruction. The color gradient indicates reconstruction quality, with yellow and red markers denoting specific evaluation points. These results collectively define a practical operating window for successful holographic SPI under varying experimental conditions.

\section{Discussion}

We introduced a holographic SPI technique for synchrotron radiation sources by employing a 2D crystal lattice reference structure in the sample plane. The approach addresses significant limitations in imaging biomolecules due to high background and prolonged radiation exposure. Through holographic enhancement from a strongly scattering lattice, the method substantially improves the signal-to-background ratio, making the signal from the target object recoverable even in scenarios where background scattering exceeds the biomolecule signal by several orders of magnitude ($\approx$ 10$^{5}$).

Numerical simulations, using the 70S ribosome as the target object, were performed to quantify the robustness and accuracy of the introduced methodology. The lattice reference effectively generates coherent interference with the weak target object signal, allowing the extraction of high-quality diffraction data under conditions that previously rendered SPI impractical at synchrotrons. Moreover, the optimization algorithm reliably reconstructs the electron density of the target object, simultaneously retrieving unknown experimental parameters such as orientation, translation shifts within the unit cell, and incident fluence. Results from multiple independent runs underscore the consistency and reproducibility of the reconstruction outcomes, as quantified by the Fourier ring correlation (FRC), which consistently exceeds the 0.5 cutoff.

Importantly, the optimized scattering strength analysis illustrates that an appropriately selected lattice strength is crucial. While choosing stronger lattice scattering generally aids the recovery of weaker signals, optimal lattice strength becomes especially critical when sample signals are relatively weak compared to the background. This insight provides clear guidance for future experimental design, emphasizing the careful tailoring of lattice references to specific experimental conditions and biological samples. Experimentally, the fabrication of tailored 2D lattice structures, potentially employing advanced lithography or colloidal self-assembly techniques, will be a critical step in practically realizing this technique.

In conclusion, our proposed lattice-based holographic SPI framework significantly expands the feasibility of high-resolution single-particle imaging at synchrotrons. By effectively mitigating background and optimizing data acquisition, this approach could enable routine structural characterization of sensitive biological samples, opening new avenues for \emph{in situ} investigations of biomolecular dynamics under near-native conditions.

Although the current study is limited to 2D numerical simulations, our method naturally extends to 3D reconstructions. The extension would introduce additional latent parameters, including 3D orientations and positions along the beam axis, thereby increasing computational complexity. In addition, the tomographic nature of the imaging protocol, where only a slice of the reciprocal structure of the object is measured in each pattern, may lead to a higher threshold in terms of required SBR. 


\begin{backmatter}
\bmsection{Funding}
Max Planck Society.


\bmsection{Disclosures}
The author declares no conflict of interest.

\bmsection{Data Availability Statement}
The data generation and reconstruction code for the 2D simulations shown here are available at : \url{https://github.com/AyyerLab/Lattice-Ref/tree/orient}.

\end{backmatter}


\bibliography{refs_clean}

\begin{thebibliography}{10}
\newcommand{\enquote}[1]{``#1''}

\bibitem{nasir2015phylogenomic}
A.~Nasir and G.~Caetano-Anoll{\'e}s, \enquote{A phylogenomic data-driven exploration of viral origins and evolution,} {\protect\JournalTitle{Science advances}} \textbf{1}, e1500527 (2015).

\bibitem{srihari2015methods}
S.~Srihari, C.~H. Yong, A.~Patil, and L.~Wong, \enquote{Methods for protein complex prediction and their contributions towards understanding the organisation, function and dynamics of complexes,} {\protect\JournalTitle{FEBS letters}} \textbf{589}, 2590--2602 (2015).

\bibitem{dean2024illuminating}
W.~F. Dean and A.~L. Mattheyses, \enquote{Illuminating cellular architecture and dynamics with fluorescence polarization microscopy,} {\protect\JournalTitle{Journal of Cell Science}} \textbf{137} (2024).

\bibitem{ekeberg2015three}
T.~Ekeberg, M.~Svenda, C.~Abergel, \emph{et~al.}, \enquote{Three-dimensional reconstruction of the giant mimivirus particle with an x-ray free-electron laser,} {\protect\JournalTitle{Physical review letters}} \textbf{114}, 098102 (2015).

\bibitem{mall2024observation}
A.~Mall, A.~Munke, Z.~Shen, \emph{et~al.}, \enquote{Observation of aerosolization-induced morphological changes in viral capsids,} {\protect\JournalTitle{arXiv preprint arXiv:2407.11687}}  (2024).

\bibitem{ayyer20203d}
K.~Ayyer, P.~L. Xavier, J.~Bielecki, \emph{et~al.}, \enquote{3d diffractive imaging of nanoparticle ensembles using an x-ray laser,} {\protect\JournalTitle{Optica}} \textbf{8}, 15--23 (2020).

\bibitem{shen2025direct}
Z.~Shen, M.~Samoli, O.~Erdem, \emph{et~al.}, \enquote{Direct observation of the exciton polaron by serial femtosecond crystallography on single cspbbr $ \_3 $ quantum dots,} {\protect\JournalTitle{arXiv preprint arXiv:2502.02343}}  (2025).

\bibitem{seibert2011single}
M.~M. Seibert, T.~Ekeberg, F.~R. Maia, \emph{et~al.}, \enquote{Single mimivirus particles intercepted and imaged with an x-ray laser,} {\protect\JournalTitle{Nature}} \textbf{470}, 78--81 (2011).

\bibitem{spence2017xfels}
J.~Spence, \enquote{Xfels for structure and dynamics in biology,} {\protect\JournalTitle{IUCrJ}} \textbf{4}, 322--339 (2017).

\bibitem{giewekemeyer2019experimental}
K.~Giewekemeyer, A.~Aquila, N.-T. Loh, \emph{et~al.}, \enquote{Experimental 3d coherent diffractive imaging from photon-sparse random projections,} {\protect\JournalTitle{IUCrJ}} \textbf{6}, 357--365 (2019).

\bibitem{sobolev2020megahertz}
E.~Sobolev, S.~Zolotarev, K.~Giewekemeyer, \emph{et~al.}, \enquote{Megahertz single-particle imaging at the european xfel,} {\protect\JournalTitle{Communications Physics}} \textbf{3}, 97 (2020).

\bibitem{PhysRevE.80.026705}
N.-T.~D. Loh and V.~Elser, \enquote{Reconstruction algorithm for single-particle diffraction imaging experiments,} {\protect\JournalTitle{Phys. Rev. E}} \textbf{80}, 026705 (2009).

\bibitem{ayyer2016dragonfly}
K.~Ayyer, T.-Y. Lan, V.~Elser, and N.~D. Loh, \enquote{Dragonfly: an implementation of the expand--maximize--compress algorithm for single-particle imaging,} {\protect\JournalTitle{Journal of applied crystallography}} \textbf{49}, 1320--1335 (2016).

\bibitem{shen2024resolving}
Z.~Shen, P.~L. Xavier, R.~Bean, \emph{et~al.}, \enquote{Resolving nonequilibrium shape variations among millions of gold nanoparticles,} {\protect\JournalTitle{ACS nano}} \textbf{18}, 15576--15589 (2024).

\bibitem{zhuang2022unsupervised}
Y.~Zhuang, S.~Awel, A.~Barty, \emph{et~al.}, \enquote{Unsupervised learning approaches to characterizing heterogeneous samples using x-ray single-particle imaging,} {\protect\JournalTitle{IUCrJ}} \textbf{9}, 204--214 (2022).

\bibitem{dold2025melting}
S.~Dold, T.~Reichenbach, A.~Colombo, \emph{et~al.}, \enquote{Melting, bubblelike expansion, and explosion of superheated plasmonic nanoparticles,} {\protect\JournalTitle{Physical Review Letters}} \textbf{134}, 136101 (2025).

\bibitem{bielecki2019electrospray}
J.~Bielecki, M.~F. Hantke, B.~J. Daurer, \emph{et~al.}, \enquote{Electrospray sample injection for single-particle imaging with x-ray lasers,} {\protect\JournalTitle{Science Advances}} \textbf{5}, eaav8801 (2019).

\bibitem{zhao2019guide}
F.-Z. Zhao, B.~Zhang, E.-K. Yan, \emph{et~al.}, \enquote{A guide to sample delivery systems for serial crystallography,} {\protect\JournalTitle{The FEBS Journal}} \textbf{286}, 4402--4417 (2019).

\bibitem{yenupuri2024helium}
T.~V. Yenupuri, S.~Rafie-Zinedine, L.~Worbs, \emph{et~al.}, \enquote{Helium-electrospray improves sample delivery in x-ray single-particle imaging experiments,} {\protect\JournalTitle{Scientific Reports}} \textbf{14}, 4401 (2024).

\bibitem{you2024impact}
T.~You, J.~Bielecki, and F.~R. Maia, \enquote{Impact of gas background on xfel single-particle imaging,} {\protect\JournalTitle{arXiv preprint arXiv:2411.16259}}  (2024).

\bibitem{helliwell1998synchrotron}
J.~Helliwell, \enquote{Synchrotron radiation and crystallography: the first 50 years,} {\protect\JournalTitle{Acta Crystallographica Section A: Foundations of Crystallography}} \textbf{54}, 738--749 (1998).

\bibitem{neutze2012time}
R.~Neutze and K.~Moffat, \enquote{Time-resolved structural studies at synchrotrons and x-ray free electron lasers: opportunities and challenges,} {\protect\JournalTitle{Current opinion in structural biology}} \textbf{22}, 651--659 (2012).

\bibitem{khubbutdinov2019coherence}
R.~Khubbutdinov, A.~Menushenkov, and I.~Vartanyants, \enquote{Coherence properties of the high-energy fourth-generation x-ray synchrotron sources,} {\protect\JournalTitle{Synchrotron Radiation}} \textbf{26}, 1851--1862 (2019).

\bibitem{nam2016fixed}
D.~Nam, C.~Kim, Y.~Kim, \emph{et~al.}, \enquote{Fixed target single-shot imaging of nanostructures using thin solid membranes at sacla,} {\protect\JournalTitle{Journal of Physics B: Atomic, Molecular and Optical Physics}} \textbf{49}, 034008 (2016).

\bibitem{hunter2014fixed}
M.~S. Hunter, B.~Segelke, M.~Messerschmidt, \emph{et~al.}, \enquote{Fixed-target protein serial microcrystallography with an x-ray free electron laser,} {\protect\JournalTitle{Scientific reports}} \textbf{4}, 6026 (2014).

\bibitem{seuring2018femtosecond}
C.~Seuring, K.~Ayyer, E.~Filippaki, \emph{et~al.}, \enquote{Femtosecond x-ray coherent diffraction of aligned amyloid fibrils on low background graphene,} {\protect\JournalTitle{Nature communications}} \textbf{9}, 1836 (2018).

\bibitem{chushkin2025prospects}
Y.~Chushkin and F.~Zontone, \enquote{Prospects for coherent x-ray diffraction imaging at fourth-generation synchrotron sources,} {\protect\JournalTitle{IUCrJ}} \textbf{12} (2025).

\bibitem{sun2018current}
Z.~Sun, J.~Fan, H.~Li, and H.~Jiang, \enquote{Current status of single particle imaging with x-ray lasers,} {\protect\JournalTitle{Applied Sciences}} \textbf{8}, 132 (2018).

\bibitem{ayyer2020reference}
K.~Ayyer, \enquote{Reference-enhanced x-ray single-particle imaging,} {\protect\JournalTitle{Optica}} \textbf{7}, 593--601 (2020).

\bibitem{mall2023holographic}
A.~Mall and K.~Ayyer, \enquote{Holographic single-particle imaging for weakly scattering, heterogeneous nanoscale objects,} {\protect\JournalTitle{Physical Review Applied}} \textbf{19}, 054027 (2023).

\bibitem{vogel2015advances}
N.~Vogel, M.~Retsch, C.-A. Fustin, \emph{et~al.}, \enquote{Advances in colloidal assembly: the design of structure and hierarchy in two and three dimensions,} {\protect\JournalTitle{Chemical reviews}} \textbf{115}, 6265--6311 (2015).

\bibitem{ye2017monolayer}
X.~Ye, J.~Huang, Y.~Zeng, \emph{et~al.}, \enquote{Monolayer colloidal crystals by modified air-water interface self-assembly approach,} {\protect\JournalTitle{Nanomaterials}} \textbf{7}, 291 (2017).

\bibitem{crowe2021structural}
C.~Crowe-McAuliffe, V.~Murina, K.~J. Turnbull, \emph{et~al.}, \enquote{Structural basis of abcf-mediated resistance to pleuromutilin, lincosamide, and streptogramin a antibiotics in gram-positive pathogens,} {\protect\JournalTitle{Nature Communications}} \textbf{12}, 3577 (2021).

\bibitem{thibault2009probe}
P.~Thibault, M.~Dierolf, O.~Bunk, \emph{et~al.}, \enquote{Probe retrieval in ptychographic coherent diffractive imaging,} {\protect\JournalTitle{Ultramicroscopy}} \textbf{109}, 338--343 (2009).

\bibitem{marchesini2008massively}
S.~Marchesini, S.~Boutet, A.~E. Sakdinawat, \emph{et~al.}, \enquote{Massively parallel x-ray holography,} {\protect\JournalTitle{Nature photonics}} \textbf{2}, 560--563 (2008).

\bibitem{li2016small}
T.~Li, A.~J. Senesi, and B.~Lee, \enquote{Small angle x-ray scattering for nanoparticle research,} {\protect\JournalTitle{Chemical reviews}} \textbf{116}, 11128--11180 (2016).

\end{thebibliography}

\end{document}